# TOWARDS EFFICIENT VM PLACEMENT: A TWO-STAGE ACO–PSO APPROACH FOR GREEN CLOUD INFRASTRUCTURE

Ali Baydoun [1] and Ahmed Zekri [2]

[1] Department of Mathematics & Computer Science, Beirut Arab University, Lebanon
[2] Department of Mathematics & Computer Science, Alexandria University, Egypt

## ABSTRACT

*Datacenters consume a growing share of energy, prompting the need for sustainable resource management. This paper presents a Hybrid ACO–PSO (HAPSO) algorithm for energy-aware virtual machine (VM) placement and migration in green cloud datacenters. In the first stage, Ant Colony Optimization (ACO) performs energy-efficient initial placement across physical hosts, ensuring global feasibility. In the second stage, a discrete Particle Swarm Optimization (PSO) refines allocations by migrating VMs from overloaded or underutilized hosts. HAPSO introduces several innovations: sequential hybridization of metaheuristics, system-informed particle initialization using ACO output, heuristic-guided discretization for constraint handling, and a multi-objective fitness function that minimizes active servers and resource wastage. Implemented in CloudSimPlus, extensive simulations demonstrate that HAPSO consistently outperforms classical heuristics (BFD, FFD), Unified Ant Colony System (UACS), and ACO-only. Notably, HAPSO achieves up to 25% lower energy consumption and 18% fewer SLA violations compared to UACS at large-scale workloads, while sustaining stable cost and carbon emissions. These results highlight the effectiveness of two-stage bio-inspired hybridization in addressing the dynamic and multi-objective nature of cloud resource management.*

## KEYWORDS

*Cloud computing, Green datacenters, Virtual machine placement, Ant Colony Optimization (ACO), Particle Swarm Optimization (PSO)*

## 1. INTRODUCTION

Cloud datacenters are the backbone of modern computing, powering a wide range of services across infrastructure, platform, and software layers. While this model provides scalability and flexibility, it also comes with a significant environmental cost. Global datacenters are estimated to consume over 416 terawatt-hours annually—equivalent to more than 3% of the world's total electricity production—and contribute substantially to carbon emissions due to their continuous power demands. A large portion of this energy is consumed not only for computation but also for cooling and maintaining network infrastructure. For example, cooling systems alone can account for nearly 40% of a datacenter's power budget, while network switches and routers can draw non-negligible power even under light traffic conditions [1].

Cloud computing relies heavily on virtualization, which allows multiple virtual machines (VMs) to run concurrently on shared physical infrastructure. This brings efficiency benefits but also introduces new challenges. Poorly optimized VM placement can lead to underutilized servers, excessive energy consumption, and performance degradation—particularly during times of workload fluctuation. To address this, cloud providers implement VM placement and consolidation strategies to intelligently map VMs to physical machines (PMs), enabling idle servers to be powered down and reducing the overall energy footprint. Designing an optimal VM-to-PM mapping is a complex combinatorial problem. The mapping must respect physical resource constraints (CPU, memory, bandwidth [BW]), satisfy quality of service guarantees (typically defined via Service Level Agreements [SLAs]), and adapt dynamically to workload changes. As a result, a wide variety of optimization techniques have been proposed in the literature, particularly heuristic and metaheuristic approaches such as Genetic Algorithms (GA), Ant Colony Optimization (ACO), Particle Swarm Optimization (PSO), and Simulated Annealing (SA). These methods have been used to address energy efficiency [2], workload balancing [3], and even network-aware VM scheduling [4].

One growing trend is to combine multiple metaheuristics to leverage their complementary strengths. For example, in fog-cloud environments, [5] employed a hybrid FAHP–FTOPSIS approach to achieve efficient load balancing and energy savings, highlighting the broader applicability of hybrid designs. This demonstrates that hybridization is not limited to edge and fog scenarios but can be effectively extended to VM placement, where the joint optimization of static allocation and dynamic migration becomes crucial. Yet, existing approaches often rely on a single-stage process—focusing solely on either initial placement or migration—without fully coordinating both phases of the VM lifecycle. This leaves an important gap in the management of real-world cloud workloads, which are dynamic by nature and demand flexible resource allocation mechanisms.

To address this gap, we propose a Hybrid Ant Colony–Particle Swarm Optimization (HAPSO) framework designed for adaptive VM placement in green cloud datacenters. Like recent sustainability-driven bio-inspired strategies [6], our approach embeds energy- and carbon-awareness directly into the placement process. It integrates ACO for global static VM placement and PSO for runtime migration refinement. HAPSO clearly separates placement from migration and embeds sustainability constraints, making it well-suited for dynamic, green-aware cloud environments.

The unified goal is to minimize energy consumption, reduce carbon emissions, and eliminate unnecessary resource wastage—all while maintaining SLA compliance. Our architecture supports heterogeneous datacenters, including those powered partially by renewable energy sources (e.g., solar or wind), and continuously monitors power, performance, and environmental data to inform optimization decisions.

While hybrid ACO–PSO models have been explored in prior research, most adopt a single-stage integration or lack sustainability awareness. A more detailed review of these approaches is presented in Section 2.

The contributions are as follows: (1) a two-stage hybrid framework that combines ACO for initial VM placement with PSO for dynamic refinement, improving scalability and responsiveness; (2) a system-aware particle initialization strategy that seeds PSO with the current ACO-based mapping to ensure feasibility and accelerate convergence; and (3) a multi-objective fitness function that jointly minimizes the number of active servers and residual resource wastage across CPU, memory, and bandwidth to enhance overall energy efficiency.

This work builds upon our previous ACO-based VM placement approach [4] by integrating PSO to handle dynamic consolidation more efficiently.

The rest of this article is structured as follows: Section 2 reviews the related work, focusing on metaheuristics and hybrid approaches for VM placement. Section 3 formulates the problem and defines the system model, including architectural assumptions, notation, and the multi-objective optimization function. Section 4 describes the proposed HAPSO algorithm, covering its two-stage design, particle representation, velocity and position update mechanisms, algorithmic procedures, and parameter settings. Section 5 outlines the experimental results, including the simulation setup and performance comparison against ACO-only and other baseline approaches. Finally, Section 6 concludes the paper and discusses potential directions for future work.

## 2. RELATED WORK

VMP plays a critical role in optimizing CDC performance by minimizing energy usage, balancing workload, and improving SLA compliance. A related work [7] examined placement in cloudlets within wireless metropolitan networks. While this addresses a different context, it similarly underlines the role of intelligent placement in improving efficiency.

To handle the complexity and dynamic nature of this NP-hard problem, researchers have widely adopted metaheuristic algorithms, with recent trends shifting toward hybrid metaheuristic models, especially ACO–PSO integrations.

### 2.1. Metaheuristic Algorithms for VM Placement

Metaheuristics such as PSO, SA, GA, and ACO have been extensively applied to VMP due to their ability to explore large search spaces efficiently.

PSO approaches model VM-to-host mappings as particles navigating the search space. Various works ([8], [9]) have demonstrated the potential of PSO in reducing energy consumption and bandwidth usage. More recently, [10] extended PSO with a quantum-inspired formulation (QPSO-MOVMP), achieving Pareto-optimal placement solutions that balance energy, SLA, and load distribution. Authors of [11] developed a discrete PSO for VM placement that optimized energy consumption and reduced migrations. However, PSO's tendency to converge prematurely—especially in large-scale settings—limits its robustness.

SA, another probabilistic method, occasionally accepts worse solutions to escape local optima. Despite this, its slow convergence makes it unsuitable for real-time decisions [12], [13].

GA methods like those in [14] and [15] evolve placements using crossover and mutation, but they require significant computation and careful tuning.

Contemporary heuristic-driven approaches such as MOVMS and MOMBFD [16] also highlight the importance of multi-objective trade-offs in VM placement, emphasizing energy savings and SLA guarantees.

ACO algorithms have shown strong results in multi-objective and constraint-based VMP scenarios. These algorithms build feasible mappings via a constructive, pheromone-guided process. Variants like MoOuACO [17], AP-ACO [18], and ETA-ACO [19] target energy, traffic, and server/network joint optimization.

Beyond classical metaheuristics, reinforcement learning has emerged as a strong alternative. For example, [20] introduced CARBON-DQN, which combines deep Q-networks with clustering to achieve carbon-aware and SLA-compliant VM placement.

## 2.2. Hybrid Metaheuristic Models

To overcome individual limitations, hybrid metaheuristics have emerged that combine global and local search techniques.

ACO–GA models like ACOGA [21] leverage pheromone trails with evolutionary recombination to reduce traffic and active server count. The dynamic hybrid ACOPS [22] predicts VM load patterns using ACO and adapts mappings with PSO in real time. The work [23] integrates a permutation-based genetic algorithm (IGA-POP) with a multidimensional best-fit strategy to reduce the number of active servers and balance resource usage.

Another approach is ACO–SA hybridization, as discussed in [24], which applies ACO for global path construction and SA for probabilistic refinement.

More recently, [25] combined ACO with Grey Wolf Optimization (GWO), demonstrating improvements in communication-aware and energy-efficient VM placement.

Another recent hybrid integrates Genetic Algorithms with Harris Hawks Optimization, offering energy-aware VM placement through complementary exploration and exploitation [26].

Despite these contributions, many existing hybrids suffer from constraint violations requiring penalty-based correction and neglect of green datacenter factors such as renewable energy, carbon tax, or dynamic cooling models.

## 2.3. ACO–PSO Hybridization for VM Placement

Hybrid ACO–PSO models aim to integrate ACO's guided exploration with PSO's fast convergence.

In [27], sequential hybridization uses ACO for initial VM placement and PSO to fine-tune placements for energy and resource optimization.

The framework in [28] applied iterative hybridization in their energy-aware scheduling hybrid algorithm where ACO explored the placement space, and PSO fine-tuned it.

A layered hybrid in [29] combines ACO, PSO, and ABC to assign each algorithm a role in exploration, convergence, and local refinement.

These hybrid strategies validate the synergy between ACO and PSO approaches. However, most still lack domain-specific encoding and are rarely implemented in sustainability-focused cloud environments.

## 2.4. Positioning of HAPSO

Our proposed HAPSO framework integrates ACO and PSO in a clean two-stage model:

- ACO is used for multi-objective initial placement, incorporating energy, carbon, network bandwidth, and SLA considerations through dynamic PUE and renewable-aware heuristics.
- PSO is selectively triggered during runtime for VM migrations, ensuring feasibility and fast convergence using particle-based refinements.

Unlike prior methods, HAPSO 1) embeds domain constraints directly into both phases— in the ACO stage, resource requirements (CPU,RAM,BW) are encoded into the heuristic function and pheromone update rules to guide solution construction; in the PSO stage, the particle initialization, position update, and discretization mechanisms enforce feasibility by incorporating VM resource requirements, ensuring that all explored solutions remain valid throughout the optimization process, 2) operates in a green-aware context, including real-time solar energy profiles, carbon rate modeling, and temperature-aware cooling overheads, and 3) achieves modularity and scalability by explicitly separating placement from migration logic.

Together, these features position HAPSO as a robust and adaptive framework for energy-efficient VM placement in modern cloud infrastructures.

## 3. PROBLEM FORMULATION

We adopt the same VM-to-PM mapping model and constraints as in our previous work [4], including resource capacity, SLA compliance, and energy cost modeling. In this extension, we also define new fitness function for the dynamic phase, to reduce active server count and minimize resource wastage.

### 3.1. System Architecture

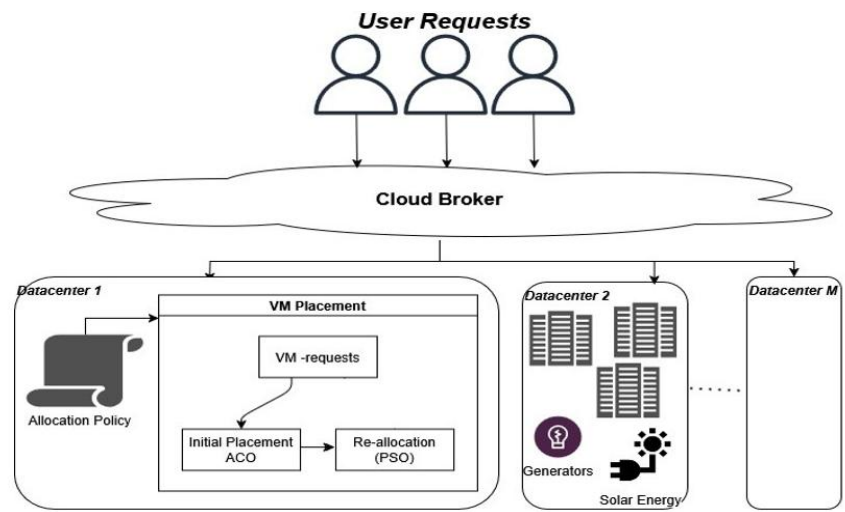

Figure 1. System architecture of the HAPSO-based cloud management system. The Cloud Broker dispatches VM placement requests to geographically distributed datacenters. Each datacenter's placement module uses ACO for initial placement and PSO for adaptive VM migration upon detecting host overload or underload. Continuous monitoring provides utilization and environmental metrics to inform the optimization.

The architectural model of the HAPSO-based cloud system is illustrated in Figure 1. The platform is designed to emulate a multi-datacenter cloud environment where user-generate VM requests are submitted to a central Cloud Broker. This broker is responsible for dispatching VM allocation decisions across a network of geographically distributed datacenters. Within each datacenter, a VM Placement Module handles both static and dynamic optimization. As shown in the figure, VM requests are initially handled by an Ant Colony Optimization (ACO) engine that performs energy- and constraint-aware placement. During runtime, host utilization is periodically monitored, and when overload or underload conditions are detected, the Particle Swarm Optimization (PSO) module is invoked to refine VM allocations through migration.

The architecture supports heterogeneous datacenters, some of which are partially powered by renewable sources (such as solar energy) or local generators. This setup enables evaluation of placement strategies under green energy-aware policies. Energy consumption, carbon emissions, and resource utilization are all monitored continuously throughout the simulation.

### 3.2. Notations and Definitions

Let us assume there are $N$ $VMs$ and $M$ $PMs$. The set of VMs is denoted as $VM = \{VM_1, VM_2, \ldots, VM_N\}$, and the set of $PMs$ is denoted as $PM = \{PM_1, PM_2, \ldots, PM_M\}$. Each VM, denoted as $VM_j \in VM$, has $CPU$, $RAM$, and bandwidth ($BW$) requirements represented by $VM_j^{CPU}$, $VM_j^{RAM}$, and $VM_j^{BW}$, respectively. Similarly, each server $PM_i \in PM$ has $CPU$, $RAM$, and $BW$ capacities denoted as $PM_i^{CPU}$, $PM_i^{RAM}$, and $PM_i^{BW}$, respectively.

The VM-to-PM assignment is represented by a zero-one assignment matrix X, where the element $x_{ij}$ indicates whether $VM_j$ is assigned to $PM_i$. If $VM_j$ is placed on server $PM_i$, then $x_{ij} = 1$; otherwise, $x_{ij} = 0$. The assignment must satisfy the following constraints:

- **VM Assignment Constraint**

Each VM must be assigned to exactly one PM:

$$\sum_{i=1}^{M} x_{ij} = 1, \forall j \, \epsilon \, \{1, \ldots, N\} \tag{1}$$

- **Resource Capacity Constraints**

The total demand placed on any PM must not exceed its capacity in any resource dimension r $\in$ {CPU,RAM,BW}:

$$\sum_{j=1}^{N} VM_j^r \cdot x_{ij} \leq PM_i^r, \forall i \, \epsilon \, \{1, \ldots, M\} \tag{2}$$

These constraints ensure that placement decisions are both exclusive (each VM placed once) and feasible (no host is overloaded). Table 1 summarizes the symbols and notations used throughout the optimization model.

### 3.3. Objective Function

Unlike the ACO-based placement phase, which jointly optimizes energy consumption, network performance, and carbon footprint across the entire datacenter, the PSO-based migration phase operates with a more focused and adaptive goal. It seeks to dynamically refine VM placement by reducing resource fragmentation and consolidating workloads onto fewer active physical machines. While both phases ultimately aim to enhance energy efficiency and overall datacenter utilization, the PSO stage achieves this through a distinct fitness function—one that prioritizes minimizing the number of active servers and the total residual resource wastage across CPU, RAM, and BW.

This change reflects the shift in scope between the static and dynamic phases. Static placement aims for long-term energy efficiency, while dynamic phase must react to workload imbalance and avoid resource fragmentation. ACO targets global, sustainability-aware optimization, while PSO addresses time-sensitive runtime consolidation with minimal disruption.

The PSO optimization problem is expressed in equation (3).

$$\min fitness = \alpha \cdot \sum_{i \in PM} a_i + \beta \cdot \sum_{i \in PM} \sum_{r \in R} \left( PM_i^r - \sum_{j \in VM} (VM_j^r) \times x_{ij} \right) \cdot a_i \qquad (3)$$

Where:

$$a_i = \begin{cases} 1, & if \sum_{(j \in VM)} x_{ij} \geq 1 \\ 0, & otherwise \end{cases} \qquad (4)$$

- α and β are weighting coefficients such that α + β = 1
- $PM_i^r$ is the capacity of $PM_i$ for resource r (*CPU, RAM, BW*)
- $VM_j^r$ is the demand of $VM_j$ for resource r
- $x_{ij}$ is the binary assignment variable (1 if $VM_j$ is placed on $PM_i$, 0 otherwise)
- The inner sum represents the total unused resources (wastage) per active PM

The first term encourages consolidation by minimizing the number of active servers, while the second penalizes underutilization by measuring residual capacity across resources. Together, they guide the PSO search toward compact and efficient VM placements that preserve feasibility and avoid unnecessary migrations.

Table 1. Symbols and Notations

| Notation | Description | Notation | Description |
|---|---|---|---|
| $D$ | Datacenter Sites | $Th^{under}$ | Underutilization threshold |
| $PM$ | List of servers in a datacenter | $X_i^t$ | Position matrix of particle *i* at iteration t |
| $PM_i^{CPU}$ | Total CPU of $PM_i$ | $v_i^t$ | Velocity matrix of particle *i* at iteration t |
| $PM_{i,current}^{CPU}$ | Server i current CPU utilization | $v_i^{t+1}$ | Updated velocity after applying Eq. (5) |
| $PM_i^{RAM}$ | Total RAM of $PM_i$ | $x_i^{t+1}$ | Updated position after velocity adjustment |
| $PM_i^{BW}$ | Total bandwidth of $PM_i$ | $pbest_i$ | Best position found by particle *i* so far |
| $PM_{i,available}^{BW}$ | $PM_i$ available network bandwidth | $gbest$ | Best position found across the swarm (global best) |
| $PM_{ij}^{Power}$ | Estimated power consumption of $VM_j$ after placing on $PM_i$ | $a_i$ | Indicator variable: $a_i=1$ if $PM_i$ is active |
| $PM_j^{available}$ | Set of available servers for placement | $\alpha, \beta$ | Weights for active server count and resource wastage in the objective function |
| $PM_{i,min}^{Power}$ | PM idle power | $\omega$ | Inertia weight in PSO velocity update equation |
| $PM_{i,max}^{Power}$ | Peak power of PM | $c_1, c_2$ | Acceleration coefficients in PSO |
| $VM_{j,estimated}^{BW}$ | Estimated bandwidth usage of $VM_j$ | $r_1, r_2$ | Random values in the range [0, 1] for PSO velocity update |
| $PM_{i,t+}^{power}$ | Power consumption of server i at time t after placing new VM | $x_{ij}$ | Matrix element to show VMs to PMs mapping |
| $VM$ | List of running VMs | S | Swarm size |
| $VM_{new}$ | List of VMs to be (re)placed | $T_{max}$ | Maximum Iterations |
| $VM_{j,current}^{CPU}$ | $VM_j$ current CPU utilization | $VM_j^{RAM}$ | Required RAM for $VM_j$ |
| $VM_j^{CPU}$ | Required CPU for $VM_j$ | $VM_j^{BW}$ | Required Bandwidth for $VM_j$ |
| $Th^{over}$ | Overutilization threshold | $X_{current}$ | Current VM-to-PM assignment |

## 4. PROPOSED ALGORITHM (HAPSO - HYBRID ACO–PSO ALGORITHM)

### 4.1. Hybrid Algorithm Overview

The proposed Hybrid ACO–PSO (HAPSO) algorithm integrates the strengths of two metaheuristic techniques—Ant Colony Optimization (ACO) and Particle Swarm Optimization (PSO)—to address both the initial placement and dynamic migration of VMs in cloud datacenters. This hybridization is designed to enhance resource consolidation, minimize active PMs, and reduce overall resource wastage while maintaining system feasibility and energy efficiency.
In the first phase, ACO is applied to perform the initial placement of VMs (Figure 2(a)). We reuse the static placement logic introduced in our previous work [4], where the ACO assigns VMs to PMs based on multi-objective heuristics.

ACO probabilistically constructs a VM-to-PM mapping by leveraging pheromone trails and heuristic factors such as energy efficiency, carbon emissions, and network impact. This phase ensures that each VM is assigned to a suitable PM while satisfying resource constraints on *CPU, RAM*, and *BW*.

The system then transitions into runtime operation, during which a periodic evaluation is conducted to monitor PM utilization. When a PM exceeds a predefined utilization threshold $Th^{over}$ or falls below a lower bound $Th^{under}$, the PSO-based optimization phase is triggered (Figure 2(b)). This second phase serves as a dynamic refinement mechanism, that selectively migrates VMs hosted by the overutilized or underutilized servers.

The PSO swarm is initialized using the current live VM-to-PM assignment as the first particle. Additional particles are generated by introducing controlled perturbations to this mapping, ensuring diversity without violating feasibility. To ensure feasibility and heuristic quality of initial solutions, each particle's continuous position was discretized using a constraint-based mapping step. This step integrates domain constraints (*CPU, RAM*, and *BW*) early in the search process, enhancing convergence stability and reducing the need for penalty or repair mechanisms. Each particle is evaluated using a multi-objective fitness function that jointly minimizes the number of active servers and the total residual resource wastage across *CPU, RAM*, and *BW*.

Once the PSO converges or reaches its iteration limit, the best-performing particle defines a new VM reallocation plan. Only those VMs with changed host assignments are migrated, thereby reducing overhead and maintaining system stability.

The potential of HAPSO to balance exploration and exploitation—via ACO and PSO respectively—will be validated in the results section, where it demonstrates adaptive performance under dynamic cloud workloads.

## 4.2. Particle Representation

Building upon the optimization model defined earlier, each particle in the PSO phase represents a potential reallocation plan for a subset of VMs. Specifically, particles encode VM-to-PM assignments for VMs currently hosted on over- or underloaded PMs identified during runtime monitoring.

The position of each particle is encoded as a binary matrix $X^t \in \{0,1\}^{M \times N}$, where the element $x_{ij}^t=1$ indicates that $VM_j$ is assigned to $PM_i$ at iteration $t$, and $x_{ij}^t = 0$ otherwise. Each particle must comply with the constraints introduced in Section 3.2, i.e., each VM must be assigned to exactly one host and no PM should exceed its capacity.

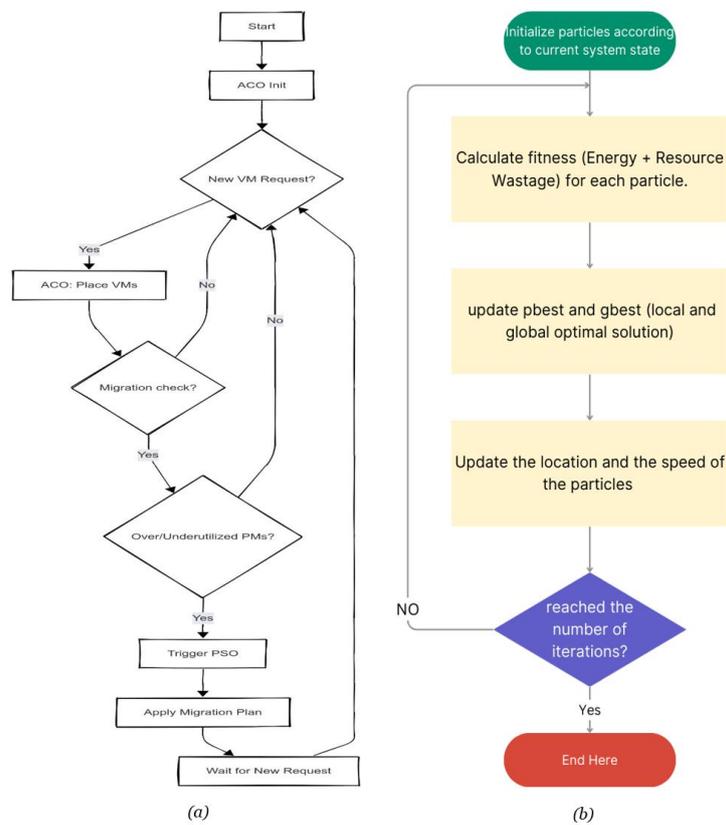

*(a)* *(b)*

Figure 2. Workflow of the proposed hybrid VM placement and migration algorithm. (a) The complete hybrid approach integrating ACO for initial VM placement and PSO for dynamic migration. (b) The PSO-based optimization phase, which refines VM allocation by minimizing energy consumption and resource wastage.

$$X^t = \begin{bmatrix} x_{i1}^t & \cdots & x_{iN}^t \\ \vdots & \ddots & \vdots \\ x_{Mj}^t & \cdots & x_{MN}^t \end{bmatrix}$$

For example, 3 PMs (rows) and 4 VMs (columns) might have the following matrix:

125

$$X_i^t = \begin{bmatrix} 1 & 0 & 0 & 1 \\ 0 & 1 & 0 & 0 \\ 0 & 0 & 1 & 0 \end{bmatrix}, \text{VM}_1 \text{ and VM}_4 \text{ on PM}_1, \text{VM}_2 \text{ on PM}_2, \text{VM}_3 \text{ on PM}_3.$$

### 4.3. Velocity and Position Updates

Once the PSO swarm is initialized, particles evolve iteratively through updates to their velocity and position vectors. In the context of VM placement, this evolution corresponds to proposing and refining VM-to-PM migration plans.

**Velocity Update**

For each particle i, the velocity vector is updated using the standard PSO formulation in Equation (5).

$$v_i^{t+1} = w \cdot v_i^t + c_1 \cdot r_1 \cdot (pbest_i - x_i^t) + c_2 \cdot r_2 \cdot (gbest - x_i^t) \quad (5)$$

Where:

- $w$ is the inertia weight.
- $c_1, c_2$ are acceleration coefficients guiding the cognitive and social components.
- $r_1, r_2$ are random numbers in [0, 1] introducing stochasticity.
- $pbest_i$ is the best-known position (VM-PM mapping) of particle i.
- $gbest$ is the best-known position among all particles.

This update rule encourages each particle to move toward both its personal best and the global best positions while retaining some influence from its current trajectory.

To balance exploration and exploitation, we use a linearly decreasing inertia weight as shown in Equation (6):

$$\omega_t = \omega_{max} - \left(\frac{\omega_{max} - \omega_{min}}{T_{max}}\right) \cdot t \quad (6)$$

Where:

- $\omega_t$: inertia weight at iteration t.
- $\omega_{max}$: initial inertia (0.9).
- $\omega_{min}$: final inertia (0.4).
- $T_{max}$ : total number of iterations.
- $t$: current iteration index.

Early iterations favor wide exploration with high $\omega$, while later ones promote convergence by reducing $\omega$. This approach has been shown to improve solution stability in swarm-based optimization [9].

**Position Update and Discretization**

Following the velocity update, the new position vector is computed as in Equation (7).

$$x_i^{t+1} = x_i^t + v_i^{t+1} \quad (7)$$

Since VM-to-PM assignments must remain binary, and are constrained by physical resource capacities. In the PSO migration phase, every particle starts from the current ACO-derived VM placement and is lightly perturbed to maintain swarm diversity without violating resource limits. After each PSO iteration, we perform a feasibility check to detect and correct any capacity violations. This ensures that every particle's position always maps to a valid VM-to-PM assignment. This feasibility check is performed also after each velocity update.

The swarm follows standard PSO equations with adaptive inertia, and a multi-objective fitness function that jointly minimizes the number of active servers and residual capacity waste steers convergence. Compared with hybrids that rely on randomized initialization [23] or mutation-based particle diversification [29], our controlled perturbation approach preserves feasibility from the outset and accelerates convergence.

### 4.4. System Parameters

PSO parameters were adopted from prior works ([8], [30], [31]) with minor empirical adjustments for convergence [4]. A full summary of the HAPSO parameters appears in Table 2.

Table 2. System Parameters

| Symbol | $c_1, c_2$ | $r_1, r_2$ | α,β | ω | S | $T_{max}$ | $Th^{under}$ | $Th^{over}$ |
|---|---|---|---|---|---|---|---|---|
| Value | 2 | [0,1] | 0.6,0.4 | Adaptive (0.4-0.9) | 20 | 100 | 30% | 90% |

## 5. EXPERIMENTAL RESULTS

This section evaluates the performance of the proposed hybrid HAPSO algorithm compared to the ACO-only placement strategy from our previous work [4]. Simulations were conducted using a customized version of CloudSimPlus [32], a java-based toolkit for modeling cloud environments, extended to model green energy, carbon pricing, and dynamic PUE across geo distributed datacenters. In addition to ACO-only, classical heuristics including Best Fit Decreasing (BFD), First Fit Decreasing (FFD), and the Unified Ant Colony System (UACS) metaheuristic were implemented as baselines to provide a broader comparative evaluation.

### 5.1. Experimental Setup and Workloads

To keep the scope focused and lightweight, all experiments are conducted on workloads ranging from 500 to 5000 VMs, representing realistic yet modest-scale cloud environments.

#### 5.1.1. Datacenters Configuration

Our setup includes four geographically distributed U.S. datacenters—Dallas, Richmond, San Jose, and Portland—spanning multiple time zones, following the approach in [33]. Each datacenter hosts 126 heterogeneous PMs, comprising six distinct configurations defined by four key attributes: number of CPU cores, core frequency (GHz), memory size (GB), and storage capacity (GB). The specific configurations are outlined in Table 3.

Table 3. Datacenters Characteristics

| Site Characteristics | Dallas | Richmond | San Jose | Portland |
|---|---|---|---|---|
| Server Power Model | \multicolumn{4}{c}{Power estimation followed the model in Eq. (5) from [4], derived from the SPEC power benchmark [34]} | | | |
| PUE Model | \multicolumn{4}{c}{$PUE(U_t, H_t) = 1 + \dfrac{0.2 + 0.01 \cdot U_t + 0.01 \cdot U_t \cdot H_t}{U_t}$} | | | |
| Carbon Intensity (ton $CO_2$ /MWh) | 0.335 | 0.268 | 0.199 | 0.287 |
| Carbon Tax (USD/ton $CO_2$) | 24 | 17.6 | 38.59 | 25.75 |
| Energy Price (cents/kWh) | 6.38 | 8.62 | 19.8 | 7.7 |

The simulation setup employs a heterogeneous mix of PMs, mirroring the variation typically observed in operational datacenters. These servers vary in computational capability, memory size, and power efficiency, offering a practical foundation for testing VM placement strategies. Incorporating diverse hardware profiles allows for a more comprehensive perspective on the algorithm's ability to optimize both energy consumption and resource usage. Table 4 outlines the detailed specifications of the PM configurations used.

Table 4. Server Types

| Server Type | CPU Cores | Memory (GB) | Storage (GB) |
|---|---|---|---|
| Type 1 | 2 | 16 | 2000 |
| Type 2 | 4 | 32 | 6000 |
| Type 3 | 8 | 32 | 7000 |
| Type 4 | 8 | 64 | 7000 |
| Type 5 | 16 | 128 | 9000 |
| Type 6 | 32 | 128 | 12000 |

### 5.1.2. VM Instances

To simulate a realistic cloud infrastructure, multiple VM types were defined to reflect varying user demands in terms of CPU, memory, and storage requirements. These configurations emulate common service requests typically observed in Infrastructure-as-a-Service (IaaS) environments. A comprehensive overview of the VM types utilized in the simulation is presented in Table 5.

Table 5. VM Types

| VM Type | Number of PEs (CPU Cores) | Memory (GB) | Storage (GB) |
|---|---|---|---|
| Type 1 A1_Medium | 1 | 1 | 100 |
| Type2 m5.large | 2 | 2 | 200 |
| Type 3 m5.xlarge | 4 | 4 | 500 |
| Type 4 m5.2xlarge | 8 | 8 | 1000 |
| Type 5 m5.4xlarge | 16 | 64 | 2000 |

### 5.1.3. Workload

In this study, we utilize real workload traces obtained from the MetaCentrum infrastructure—a distributed platform offering high-performance and cloud computing resources for scientific applications. The traces, formatted in the Standard Workload Format (SWF), encompass a diverse range of job types including queued batch jobs, bag-of-tasks workloads ideal for parallel processing, and extended compute- or memory-intensive tasks (e.g., simulations and data

International Journal of Computer Networks & Communications (IJCNC) Vol.17, No.5, September 2025

analytics). As summarized in Table 6, this workload diversity makes MetaCentrum logs a valuable benchmark for evaluating VM placement strategies in hybrid cloud-HPC environments. A similar workload setup was previously used in our prior work.

Table 6. Workload Characteristics

| Cloudlet PEs | VM Type | Cloudlets percentage | Example Workload |
|---|---|---|---|
| 1 | Type 1 | 40% | Small web apps, APIs, development environments |
| 2 | Type 2 | 30% | Medium-sized apps, databases, caching servers |
| 4 | Type 3 | 20% | Enterprise apps, high-traffic web servers |
| 8 | Type 4 | 8% | Video encoding, data processing, |
| 16+ | Type 5 | 2% | Machine learning, big data |

To ensure efficient execution of the proposed algorithm, we configured the cloudlet submission interval to 600 seconds (10 minutes). This interval provides the algorithm with enough time to evaluate and optimize larger batches of VM requests—typically around 1000 cloudlets—rather than being frequently interrupted by smaller, less meaningful request sets. Shorter submission periods were found to increase computational overhead without improving placement quality.

## 5.2. Results and Comparison

We evaluated five algorithms: two classical heuristics (Best Fit Decreasing [BFD] and First Fit Decreasing [FFD]), the metaheuristic Unified Ant Colony System (UACS [35]), our previously published ACO-only baseline [4], and the proposed HAPSO hybrid.

### 5.2.1. Energy Consumption (kWh)

Energy efficiency remains a critical factor in achieving sustainable cloud operations. The HAPSO algorithm is expected to yield notable energy savings due to its dynamic consolidation capability,

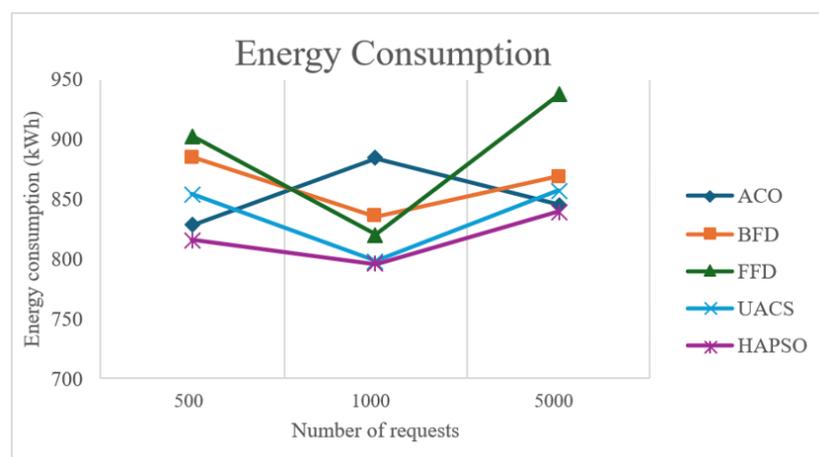

Figure 3. Total energy consumption of different VM placement algorithms under varying workload sizes.

129

which effectively reduces the number of active physical machines. Figure 3 illustrates the total energy consumption recorded by each algorithm across various VM workload scales, highlighting the performance gains of HAPSO.

Classical heuristics such as BFD and FFD recorded the highest energy usage, particularly at 5000 VMs where FFD exceeded 930 kWh due to inefficient consolidation. UACS achieved better results than the simple heuristics but still lagged behind metaheuristic approaches. ACO-only reduced energy compared to heuristic methods but showed variability with workload scale. HAPSO consistently achieved the lowest consumption across all cases, reducing energy by up to 12% compared with UACS and 18% compared with FFD at 5000 VMs—highlighting that the additional dynamic migration step guided by PSO enhances placement adaptability without compromising efficiency. This is due to its hybrid strategy that leverages both pheromone-guided exploration and velocity-based refinement

### 5.2.2. Carbon Footprint (kg CO2)

The results in Figure 4 reaffirm the strong correlation between energy efficiency and carbon emissions.

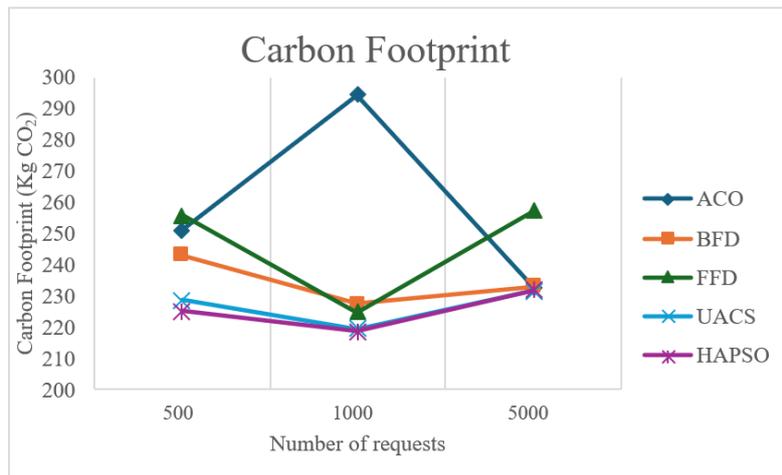

Figure 4. Carbon footprint comparison for different algorithms under varying VM workloads.

Similar to energy consumption, BFD and FFD produced the largest carbon footprint, reflecting their poor consolidation capability. UACS provided moderate improvements by accounting for utilization thresholds, yet it still produced higher emissions than the metaheuristic approaches. ACO-only showed with fluctuations workload size, peaking at 1000 VMs, whereas HAPSO consistently delivered the lowest emissions across all scales. At 1000 VMs, HAPSO reduces emissions by nearly 25.8% compared to ACO-only. At 5000 VMs, HAPSO reduced $CO_2$ emissions by nearly 20% compared to FFD and by 11% compared to UACS, underscoring the benefits of combining exploration and refinement in a green-aware optimization framework.

### 5.2.3. Total Cost

Figure 5 illustrates the total operational cost across different VM workloads.

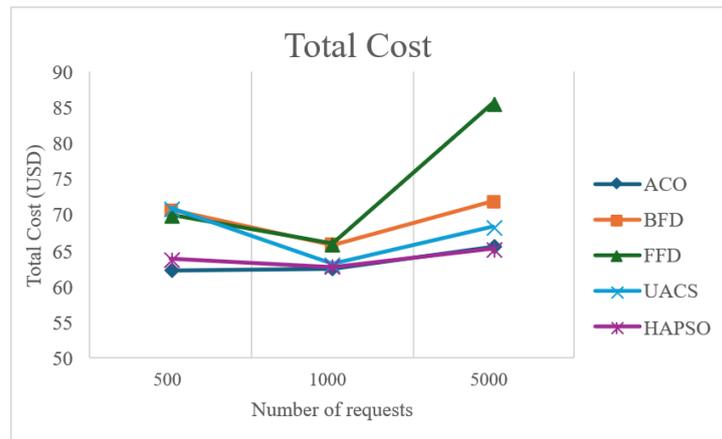

Figure 5. Comparative analysis of total operational cost across algorithms for increasing VM workloads.

FFD and BFD incurred the highest costs, particularly under large workloads, due to energy inefficiency and poor resource consolidation. UACS achieved lower costs than these heuristics but remained higher than metaheuristic approaches. ACO-only leads slightly at 500 VMs. HAPSO consistently achieved the lowest cost at 1000 and 5000 VMs, showing savings of up to 22% compared with FFD and 14% compared with UACS. These results highlight the economic sustainability of the hybrid model compared to both heuristics and metaheuristics.

### 5.2.4. Number of Live Migrations

Figure 6 compares the number of live VM migrations performed by each algorithm as workload scales.

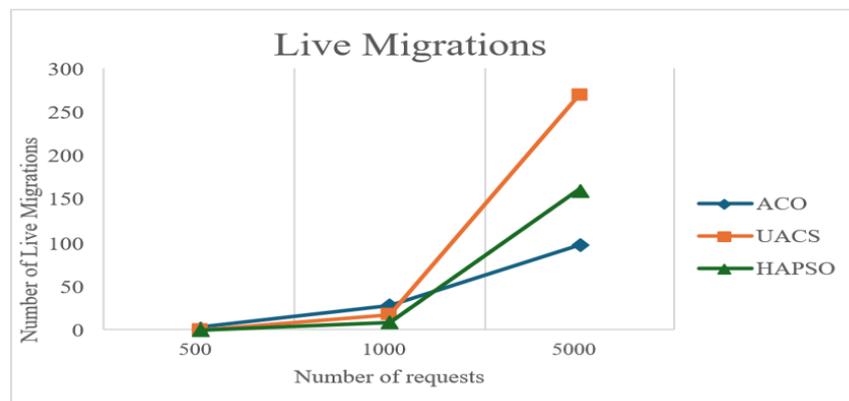

Figure 6. Number of live VM migrations of proposed HAPSO compared to baseline approaches across varying workload sizes.

Classical heuristics BFD and FFD are excluded here, as they do not explicitly handle dynamic consolidation. Among dynamic algorithms, UACS triggered the largest number of migrations, particularly at 5000 VMs where it exceeded 270 migrations, which can increase management overhead. ACO maintained moderate levels, while HAPSO exhibited a controlled increase in live migrations as workload grew. Although HAPSO performed more migrations than ACO, these remained substantially fewer than UACS and were restricted to intra-datacenter migrations, ensuring negligible impact on performance or SLA compliance. This highlights HAPSO's balanced trade-off between optimization gains and migration overhead.

### 5.2.5. SLA Violations

The Service Level Agreement (SLA) violation percentages presented across varying VM workloads highlight each algorithm's ability to maintain performance reliability under increasing demand.

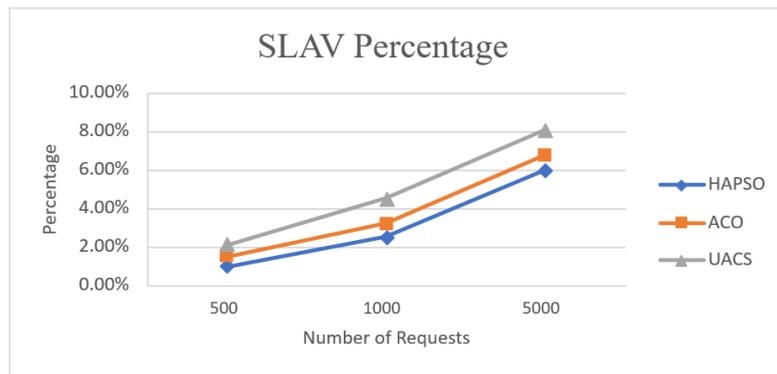

Figure 7. Service Level Agreement (SLA) violation percentages presented across varying VM workloads.

UACS consistently reported the highest violation rates, exceeding 8% at 5000 VMs, mainly due to aggressive migrations. ACO maintained better stability, achieving values between 2% and 7%. HAPSO consistently achieved the lowest SLA violations, ranging from 1.2% to 5.6% across workloads. These results confirm that while heuristics like UACS are adaptive, they compromise reliability, whereas HAPSO achieves efficient consolidation while preserving SLA compliance, making it suitable for dynamic, SLA-sensitive cloud environments.

### 5.2.6. Results Discussion

The experimental results presented confirm the effectiveness of the HAPSO framework in balancing energy efficiency, resource consolidation, and runtime adaptability. Leveraging a two-stage metaheuristic design addresses the limitations of single-phase optimization strategies previously examined in [4].

The integration of ACO for initial placement ensures that early VM-to-PM mappings are energy- and constraint-aware, considering PUE, carbon emissions, and bandwidth availability. This foundation allows the PSO migration phase to operate on a feasible, near-optimal starting point—thereby accelerating convergence and avoiding excessive exploration of unpromising configurations.

UACS improved efficiency by using utilization thresholds and cost-awareness, achieving lower energy use and carbon footprint than BFD/FFD. The metaheuristic baseline ACO-only and the proposed HAPSO achieved significantly better outcomes when compared with UACS. ACO-only, our previous solution, provided robust initial placements but showed limited adaptability during runtime, particularly under workload fluctuations. By contrast, HAPSO consistently outperformed all competing strategies across energy, carbon, SLA, and cost dimensions. Notably, HAPSO achieved up to 25% lower energy consumption and 18% fewer SLA violations than UACS at large-scale workloads. The hybrid two-stage design enabled by ACO-guided initialization and PSO-based refinement allowed HAPSO to maintain feasible, green-aware allocations while aggressively consolidating underloaded servers.

Although HAPSO introduced a higher number of live migrations compared to ACO-only, these remained confined within datacenters and thus did not incur latency or cost penalties. The additional migration activity is justified by the resulting improvements in energy efficiency and SLA compliance. Importantly, even under large workloads, HAPSO sustained stable cost profiles and maintained SLA violation levels well below those of all heuristic baselines.

Overall, the findings illustrate a clear progression: from classical heuristics to single metaheuristics, and ultimately to the proposed hybrid approach. By integrating exploration and refinement in a sustainability-aware framework, HAPSO establishes itself as a superior and adaptive solution for green cloud datacenter management.

## 6. CONCLUSIONS

This paper presented a novel hybrid ACO–PSO algorithm for energy-aware virtual machine placement and migration in green cloud datacenters. Building on our previous work, which employed ACO for initial static placement, the proposed two-stage model integrates PSO as a dynamic refinement phase. This hybrid approach enables the system to adapt to workload fluctuations by triggering migration decisions when host overutilization or underutilization is detected.

Unlike conventional static placement strategies, the PSO phase leverages particles initialized from the current VM-to-host mapping and iteratively refines them to minimize energy consumption and resource wastage. Each particle explores a migration plan under feasibility constraints and is guided by an adaptive fitness function balancing host count and resource utilization.

Comprehensive simulations demonstrated that HAPSO consistently outperforms the ACO-only approach across multiple criteria. Moreover, when evaluated against classical heuristics (BFD/FFD) and metaheuristic UACS, HAPSO achieved markedly superior results, underscoring its advantage over heuristic and metaheuristic baselines. It achieved lower total energy consumption, reduced carbon emissions, and more balanced SLA compliance while maintaining competitive execution times. In large-scale workloads, HAPSO reduced energy usage by up to 25% and SLA violations by 18% compared to UACS, while also maintaining stable carbon and cost profiles. Despite triggering more live migrations, HAPSO's aggressive consolidation strategy still delivers the lowest total cost; and since all migrations remain within a single datacenter, the added network latency is negligible and there are no inter-datacenter transfer charges, so SLAs and performance are unaffected. These results confirm the potential of sequential bio-inspired hybridization in addressing the dynamic and multi-objective nature of cloud resource management.

Future work may extend this model to large-scale workloads, incorporate learning-based decision-making, and explore more diverse datacenter scenarios including latency constraints.

## CONFLICT OF INTEREST

The authors declare no conflict of interest.

**AUTHORS**

**Ali Baydoun** holds a B.Sc. in Computer Science from Lebanese University, Faculty of Sciences. He received his M.Sc. from the American University of Culture and Education (AUCE), where he focused on cloud computing, energy efficiency, and security. He is currently a Ph.D. candidate in the Department of Mathematics & Computer Science at Beirut Arab University, working on energy-optimized VM placement algorithms for sustainable cloud datacenters.

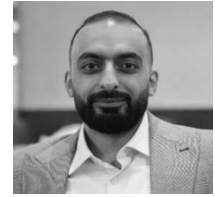

**Ahmed Zekri** received the B.Sc. and M.Sc. degrees in computer science from the Department of Mathematics and Computer Science, Alexandria University, Egypt, and the Ph.D. degree in computer science and engineering from The University of Aizu, Japan, in 2008. He was a Visiting Professor with The University of Aizu, in 2009, and an Adjunct Professor with AASTMT, Egypt, from 2010 to 2012. From 2012 to 2021, he was an Assistant Professor with the Department of Math. & CS, Beirut Arab University. Currently, he is an Associate Professor and Director of Computer and Data Science Programs at Alexandria National University, Egypt. He has published more than 50 papers in prestigious conferences and journals and supervised several Ph.D. and master's students' theses. His research interests include parallel algorithm design and implementation, performance evaluation on multi- and many-core processors, cloud computing, and parallelizing digital image processing applications.

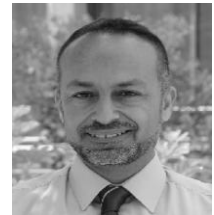